\documentclass[hyper]{JHEP3}
\skip\footins = 1\bigskipamount plus 2pt minus 4pt
\usepackage{epsfig}
\usepackage{subfigure}
\usepackage{amsmath}
\usepackage{amsfonts}
\usepackage{amssymb}
\newcommand{\be}{\begin{equation}}
\newcommand{\ee}{\end{equation}}
\newcommand{\bea}{\begin{eqnarray}}
\newcommand{\eea}{\end{eqnarray}}

\newcommand{\Slash}[1]{\ooalign{\hfil/\hfil\crcr$#1$}}

\title{Neutrino Oscillations in a Supersymmetric 
SO(10) Model with Type-III See-Saw Mechanism}

\author{Takeshi Fukuyama$^{\dagger}$, Amon Ilakovac$^{\ddagger}$, 
Tatsuru Kikuchi$^{\dagger}$ and Koichi Matsuda$^{\star}$\\
$^{\dagger}$ 
Department of Physics, Ritsumeikan University\\ 
Kusatsu, Shiga, 525-8577 Japan\\
E-mail: \email{fukuyama@se.ritsumei.ac.jp},
\email{rp009979@se.ritsumei.ac.jp}\\
$^{\ddagger}$ 
University of Zagreb, Department of Physics\\
P.O. Box 331, Bijeni\v cka cesta 32, HR-10002 Zagreb, Croatia\\
E-mail: \email{ailakov@rosalind.phy.hr}\\
$^{\star}$
Department of Physics, Osaka University\\ 
Toyonaka, Osaka, 560-0043, Japan\\
E-mail: \email{matsuda@het.phys.sci.osaka-u.ac.jp}}

\abstract{
The neutrino oscillations are studied in the framework of 
the minimal supersymmetric SO(10) model with Type-III see-saw 
mechanism by additionally introducing a number of SO(10) singlet 
neutrinos. The light Majorana neutrino mass matrix is given by 
a combination of those of the singlet neutrinos and the $SU(2)_L$ active 
neutrinos. The minimal SO(10) model gives an unambiguous Dirac 
neutrino mass matrix, which enables us to predict the masses and 
the other parameters for the singlet neutrinos. 
These predicted masses take the values accessible and testable 
by near future collider experiments under the reasonable assumptions. 
More comprehensive calculations on these parameters are also given. }

\preprint{}
\begin{document}
\section{Introduction}
As pointed out in \cite{Weinberg:1979sa}, we can construct, 
within the context of the standard model (SM), an operator 
which gives rise to the neutrino masses as 
\be
{\cal L}_{\rm eff} = \frac{1}{\Lambda} \,(\ell_L H)^{T} C^{-1} (\ell_L H)\;.
\ee
Here $\ell$, $H$ are the lepton doublet and the Higgs doublet, 
$C$ is the charge conjugation operator and $\Lambda$ is the scale 
in which something new physics appears. 
In the usual see-saw mechanism (type-I see-saw mechanism) \cite{see-saw}, 
the scale parameter $\Lambda$ is interpreted as the energy scale at which 
the right-handed neutrinos become active. 
In this paper, we explore the other possibility of type-III seesaw, 
introducing a set of singlet into the minimal supersymmetric 
standard model (MSSM). 
The motivations of this are as follows. One comes from the theoretical 
reason that string inspired $E_6$ models include SO(10) singlets 
as a matter content. 
The other does from the empirical reason that many indicate reduced 
coupling of neutrinos to the $Z^0$-boson in the framework of the SM 
or the SM with right-handed neutrinos \cite{Zeller:2001hh, Takeuchi:2002nn}. 

\section{Type-III see-saw mechanism}
We begin with reviewing the essential concept of 
the type-III see-saw mechanism proposed in the reference 
\cite{EW86, Mohapatra:1986aw, Mohapatra:1986bd, Val86, Barr:2003nn}. 
You can find a detailed study in \cite{Melo:1996ht}. 
In this model, in addition to the usual $SU(2)_L$ singlet 
$N = \nu_R$, we add a new SO(10) singlet neutrino ``$S$'', 
which has a positive lepton number (+1), 
\be
{\cal L}_Y = \int d^2 \theta \left(
Y_\nu ~\overline{N} \ell_L \,H_u 
+ Y_s ~\overline{N} S_L \,H_s
+ \mu_s ~ S_L^{T} C^{-1} S_L \right) + h.c. \;,
\label{typeIII}
\ee
where $H_u$ and $H_s$ are the $SU(2)_L$ doublet and singlet 
chiral superfields, respectively. 
This Lagrangian is written in a matrix form in the base with 
$\{\nu_L,~N,~S_L \}$ as follows: 
\be
\left(
\begin{array}{ccc}
0    & m_D^T  &  0   \\
m_D  & 0      &  M_D^T \\
0    & M_D    & \mu_s  
\end{array}
\right)\;.
\ee
After the spontaneous symmetry breaking, they give masses to 
the neutrinos as 
\bea
m_D &=& Y_\nu \left<H_u^0 \right>~,~~~M_D \ =\ Y_s \left<H_s \right>~.
\label{MD}
\eea
Note that the $\mu_s$ term in the above breaks an originally existing 
global $U(1)_L$ and $U(1)_{\cal R}$ symmetries. Thus we can naturally 
expect it as a small value compared with the electroweak scale 
even around the keV scale, 
according to the following reason: 
when the $\mu_s$ term is arisen from the VEV of a singlet 
$\mu_s = \lambda \left<S' \right>$, there appears a pseudo-NG boson, 
called Majoron $J=\Im S'$ associated with the spontaneously broken 
$U(1)_L$ symmetry. Then the keV scale lepton number violation may lead 
to an interesting signature in the neutrinoless double beta decay 
\cite{Berezhiani:1992cd} or becomes a possible candidate for 
the cold dark matter \cite{Berezinsky:1993fm}.

By integrating out $\nu_R$,
\be
\frac{\partial{{\cal L}}}{\partial{N}} = m_D \nu_L + M_D S_L =0 \;, 
\ee
we obtain 
\be
S_L = -\left(\frac{m_D}{M_D} \right) \nu_L \;.
\label{SL}
\ee
This means that the light neutrino mass eigenstate is a linear combination 
of two states $\nu_L$ and $S_L$ with the mixing angle 
$\epsilon \cong m_D/M_D$: 
\be
\nu_{\rm light} = \nu_L - \epsilon~ S_L\;.
\ee
Such an extra mixing term is interesting when we try to explain 
the ``NuTeV anomaly'' through the heavy singlet neutrino contributions 
to the neutrino--nucleon scatterings \cite{Zeller:2001hh, Takeuchi:2002nn}. 

Putting Eq.~(\ref{SL}) into Eq.~(\ref{typeIII}), 
we get the effective light neutrino mass matrix as 
\be
M_\nu =  \mu_s \left(\frac{m_D^{T} m_D}{M_D^2} \right)\;.
\ee
In general, adding three singlet neutrinos $\{S_1, S_2, S_3\}$, the 
effective light neutrino mass matrix can be written in the matrix form as 
\be
M_\nu =  \left(M_D^{-1} m_D \right)^{T} \mu_s 
\left(M_D^{-1} m_D \right)\;.
\label{typeIII}
\ee
This matrix is diagonalised by Maki-Nakagawa-Sakata (MNS) 
mixing matrix $U$ as 
\be
U^{T} M_\nu ~U = {\rm diag} (m_1,m_2,m_3)\;.
\ee
An important fact is that the new physics scale has also the 
``see-saw structure'' as 
\be
\Lambda \cong \frac{M_D^2}{\mu_s}\;.
\ee
Hence this mechanism is sometimes called as ``double see-saw'' mechanism. 
It's not the actual see-saw type but the inverse see-saw form, because 
the small lepton number violating ($\Slash{L}$) scale $\mu_s$ 
would indicate the large scale. 

Now we consider the general three generation cases. 
For simplicity, we assume that all $S_i$ have a common $\mu_s$ term. 
Then the light neutrino matrix is written as 
\be
\mu_s \, U^{T} m_D^{T} (M_D M_D^{T})^{-1} m_D U
= {\rm diag}(m_1,m_2,m_3)\;,
\ee
that is, 
\be
M_D M_D^T
= m_D U {\rm diag}\left( \frac{\mu_s}{m_1}, \frac{\mu_s}{m_2}, 
\frac{\mu_s}{m_3} \right) U^T m_D^T \;.
\ee
This symmetric combination can be diagonalised by a single unitary 
matrix ${\cal U}$ 
\be
{\cal U}^T ~ M_D M_D^T ~{\cal U}
= {\rm diag}(M_{D1}^2, M_{D2}^2, M_{D3}^2) \;.
\label{U}
\ee
Here we note that \(\cal U\) includes three mixing angles 
$\theta'_{1}$, $\theta'_{2}$, $\theta'_{3}$ and six phases 
(\(\delta\), \(\zeta_2^L\), \(\zeta_3^L\), 
\(\zeta_1^R\), \(\zeta_2^R\), \(\zeta_3^R\))
\be
{\cal U} = 
\left(
\begin{array}{ccc}
1 & 0              & 0            \\
0 & e^{i\zeta_2^L} & 0            \\
0 & 0              & e^{i\zeta_3^L}\\
\end{array}
\right)
\left(
\begin{array}{ccc}
c_{3}c_{1} & c_{3}s_{1} & s_{3}e^{-i\delta} \\
-c_{2}s_{1}-s_{2}c_{1}s_{3} e^{i\delta}
&c_{2}c_{1}-s_{2}s_{1}s_{3} e^{i\delta} 
&s_{2}c_{3} \\
s_{2}s_{1}-c_{2}c_{1}s_{3} e^{i\delta}
 & -s_{2}c_{1}-c_{2}s_{1}s_{3} e^{i\delta} 
& c_{2}c_{3} \\
\end{array}
\right)
\left(
\begin{array}{ccc}
e^{i\zeta_1^R} & 0              & 0            \\
0              & e^{i\zeta_2^R} & 0            \\
0              & 0              & e^{i\zeta_3^R}\\
\end{array}
\right),
\ee
where $s_{i}:= \sin \theta'_{i}$, $c_{i}:=\cos \theta'_{i}$. 
You should not confuse these mixing angles with those of the MNS 
mixing matrix $U$ appearing in Eq.~(\ref{MNS}). 

From this expression, we can obtain a prediction about masses and mixings 
for the heavier Dirac mass matrix $M_D$ by giving some informations about 
the light neutrino masses and mixings and the lighter Dirac mass matrix $m_D$. 

\section{Fermion masses in an SO(10) Model with a singlet}
In order to make a prediction on the second Dirac neutrino mass matrix 
$M_D$, we need an information for the Yukawa couplings of $Y_\nu$. 
In this paper, we make the minimal SO(10) model extend to add 
a number of singlet, which preserves a precise information for $m_D$. 
We begin with a review of the minimal SUSY SO(10) model proposed in 
\cite{Babu:1992ia} and recently analysed in detail in references 
\cite{Matsuda:2000zp, Matsuda:2001bg, Fukuyama:2002ch, 
Goh:2003sy, Goh:2003hf, Bajc:2002iw, Dutta:2004wv, Matsuda:2004bq}. 
Even when we concentrate our discussion on the issue of how to reproduce 
the realistic fermion mass matrices in the SO(10) model, there are lots of 
possibilities of the introduction of Higgs multiplets. 
The minimal supersymmetric SO(10) model includes only one {\bf 10} 
and one $\overline{\bf 126}$ Higgs multiplets in Yukawa couplings 
with {\bf 16} matter multiplets. Here, in addition to it, 
we introduce a number of SO(10) singlet chiral superfields ${\bf 1}$ 
as new matter multiplets
\footnote{The singlet matter multiplet may have it's origin in some $E_6$ 
representations ${\bf 27}$ or ${\bf 78}$ which are decomposed under the SO(10) 
subgroup as ${\bf 27} = {\bf 16 + 10 +1}$, 
${\bf 78} = {\bf 45 + 16 + \overline{16} + 1}$. 
In such a case, the superpotential given in Eq. (\ref{WY}) may be generated 
from the following $E_6$ invariant superpotential: 
$W_Y = Y_{1}^{ij} {\bf 27}_i {\bf 27}_j {\bf 27}_H 
+ Y_{2}^{ij} {\bf 27}_i {\bf 27}_j {\bf 351}^\prime_H 
+ Y_{3}^{ij} {\bf 27}_i {\bf 78}_j \overline{{\bf 27}}_H$. 
}. 
This additional singlet can provide a type-III see-saw mechanism as 
described in the previous section. 
In order to avoid a large triplet VEV for $\overline{\bf 126}_H$ 
unnecessary in type-III see-saw model, 
we use a $U(1)_{\cal R}$ symmetry. The corresponding $U(1)_{\cal R}$ 
charges are listed in Table 1. 
Then the relevant superpotential can be written as
\begin{table}
\begin{center}
\begin{tabular}{|c|c|c|}
\hline \hline
fields & $U(1)_{\cal R}$ charges \\
\hline
${\bf 16}_i$ & $-1$ \\
${\bf 10}_H$ & $+4$ \\
$\overline{\bf 126}_H$ & $+4$ \\
$\overline{\bf 16}_H$ & $+2$ \\
${\bf 16}_H$ & $-2$ \\
${\bf 1}_i$ & $+1$ \\
\hline \hline
\end{tabular}
\caption{$U(1)_{\cal R}$ charges of the fields relevant 
for the quark and lepton mass matrices ($R[W]=+2$). }
\end{center}
\end{table}
\be
W_Y = Y_{10}^{ij} {\bf 16}_i {\bf 16}_j {\bf 10}_H
+ Y_{126}^{ij} {\bf 16}_i {\bf 16}_j \overline{{\bf 126}}_H
+ Y_{s}^{ij} {\bf 16}_i {\bf 1}_j \overline{{\bf 16}}_H
+ \mu_s {\bf 1}_i^2 \;.
\label{WY}
\ee
At low energy after the GUT symmetry breaking, the superpotential leads to 
\begin{eqnarray}
W &=&
\left(Y_{10}^{ij} H_{10}^u + Y_{126}^{ij} H_{126}^u \right) u^c_i q_j
+
\left(Y_{10}^{ij} H_{10}^d + Y_{126}^{ij} H_{126}^d \right) d^c_i q_j
\nonumber\\
&+&
\left(Y_{10}^{ij} H_{10}^u -3  Y_{126}^{ij} H_{126}^u \right) N_i {\ell}_j
+
\left(Y_{10}^{ij} H_{10}^d -3 Y_{126}^{ij} H_{126}^d \right) e^c_i {\ell}_j
\nonumber\\
&+& Y_s^{ij} N_i S_j H_s + \mu_s S_i^2 \;, 
\end{eqnarray}
where $H_{10}$ and $H_{126}$ correspond to the Higgs doublets in ${\bf 10}_H$ 
and $\overline{{\bf 126}}_H$. That is, we have two pairs of Higgs doublets. 
In order to keep the successful gauge coupling unification, 
we suppose that one pair of Higgs doublets 
(a linear combination of $H_{10}^{u,d}$ and $H_{126}^{u,d}$) 
is light while the other pair is  heavy ($\simeq M_{\rm GUT}$). 
The light Higgs doublets are identified 
as the MSSM Higgs doublets ($H_u$ and $H_d$) and given by 
\be
H_u \ =\ \widetilde{\alpha}_u ~H_{10}^u + \widetilde{\beta}_u ~H_{126}^u \;;
\quad
H_d \ =\ \widetilde{\alpha}_d ~H_{10}^d + \widetilde{\beta}_d ~H_{126}^d \;,
\label{mix}
\ee
where $\widetilde{\alpha}_{u,d}$ and $\widetilde{\beta}_{u,d}$ denote 
elements of the unitary matrix which rotate the flavour basis in the original 
model into the SUSY mass eigenstates. Omitting the heavy Higgs mass 
eigenstates, the low energy superpotential is described by only the light 
Higgs doublets $H_u$ and $H_d$ such that 
\begin{eqnarray}
W_Y &=&
\left( \alpha^u  Y_{10}^{ij} + \beta^u  Y_{126}^{ij} \right) u^c_i q_j H_u
\ +\
\left( \alpha^d  Y_{10}^{ij} + \beta^d  Y_{126}^{ij} \right) d^c_i q_j H_d
\nonumber\\
&+&
\left( \alpha^u  Y_{10}^{ij} -3 \beta^u Y_{126}^{ij} \right) N_i \ell_j H_u
\ +\
\left( \alpha^d  Y_{10}^{ij} -3 \beta^d Y_{126}^{ij} \right) e_i^c \ell_j H_d
\nonumber\\
&+& Y_s^{ij} N_i S_j H_s \ +\ \mu_s S_i^2 \;,
\label{Yukawa3}
\end{eqnarray} 
where the formulas of the inverse unitary transformation of Eq.~(\ref{mix}), 
$H_{10}^{u,d} = \alpha^{u,d} H_{u,d} + \cdots $ and 
$H_{126}^{u,d} = \beta^{u,d} H_{u,d} + \cdots $, have been used. 
Providing the Higgs VEV's, $\langle H_u \rangle = v \sin \beta$ and 
$\langle H_d \rangle = v \cos \beta$ with $v \simeq 174$ [GeV], 
the Dirac mass matrices can be read off as 
\begin{eqnarray}
M_u &=& c_{10} M_{10} + c_{126} M_{126}, 
\nonumber \\
M_d &=& M_{10} + M_{126},   
\nonumber \\
m_D &=& c_{10} M_{10} - 3 c_{126} M_{126}, 
\nonumber \\
M_e &=& M_{10} - 3 M_{126}, 
\label{massmatrix}
\end{eqnarray} 
where $M_u$, $M_d$, $m_D$ and $M_e$ denote up-type quark, down-type quark, 
Dirac neutrino and charged-lepton mass matrices, respectively. Note that 
all the quark and lepton mass matrices are characterised by only two basic 
mass matrices, $M_{10}$ and $M_{126}$, and four complex coefficients 
$c_{10}$ and $c_{126}$. In addition to the above mass matrices the above 
model indicates the mass matrices, 
\begin{eqnarray}
M_R &=& c_R~M_{126}\;,
\nonumber\\
M_L &=& c_L~M_{126}\;,
\label{massmatrix2}
\end{eqnarray} 
together with $M_D$ given in Eq. (\ref{MD}). 
$c_R$ and $c_L$ correspond to the VEV's of 
$({\bf 10}, {\bf 1}, {\bf 3}) \subset {\overline{\bf126}}$ 
and $({\overline{\bf 10}}, {\bf 3}, {\bf 1}) \subset {\overline{\bf126}}$, 
respectively \cite{Matsuda:1999yx}. If $M_R$, $M_L$, $M_D$ terms dominate, 
they are called Type-I, Type-II, and Type-III see-saw, respectively. 
In this paper, we consider the case $c_R \ =\ c_L \ =\ 0 $, Type-III. 
Here $c_R =0$ means that the theory does not pass the Pati-Salam phase 
and is broken to the standard model directly.

The mass matrix formulas in Eq.~(\ref{massmatrix}) leads to the GUT 
relation among the quark and lepton mass matrices, 
\begin{eqnarray}
M_e = c_d \left( M_d + \kappa  M_u \right) \; , 
\label{GUTrelation} 
\end{eqnarray} 
where 
\begin{eqnarray}
c_d &=& - \frac{3 c_{10} + c_{126}}{c_{10}-c_{126}}, 
\\
\kappa &=& - \frac{4}{3 c_{10} + c_{126}}. 
\end{eqnarray} 
Without loss of generality, we can take the basis where $M_u$ is real 
and diagonal, $M_u = D_u$. Since $M_d$ is the symmetric matrix, it is 
described as $M_d = V_{\mathrm{CKM}}^* \,D_d \,V_{\mathrm{CKM}}^\dagger$ 
by using the CKM matrix $V_{\mathrm{CKM}}$ and the real diagonal mass matrix 
$D_d$. 
Considering the basis-independent quantities, $\mathrm{tr}[M_e^\dagger M_e ]$, 
$\mathrm{tr} [(M_e^\dagger M_e)^2 ]$ and $\mathrm{det} [M_e^\dagger M_e ]$, 
and eliminating $|c_d|$, we obtain two independent equations,  
\begin{eqnarray}
\left(
\frac{\mathrm{tr} [\widetilde{M_e}^\dagger \widetilde{M_e} ]}
{m_e^2 + m_{\mu}^2 + m_{\tau}^2} \right)^2
&=& 
\frac{\mathrm{tr} [( \widetilde{M_e}^\dagger \widetilde{M_e} )^2 ]}
{m_e^4 + m_{\mu}^4 + m_{\tau}^4},
\label{cond1} \\ 
\left( \frac{\mathrm{tr} [\widetilde{M_e}^\dagger \widetilde{M_e} ]}
{m_e^2 + m_{\mu}^2 + m_{\tau}^2} \right)^3
&=&
\frac{\mathrm{det} [\widetilde{M_e}^\dagger \widetilde{M_e} ]}
{m_e^2 \; m_\mu^2 \; m_\tau^2},
\label{cond2} 
\end{eqnarray}
where $\widetilde{M_e} \equiv 
V_{\mathrm{CKM}}^* \, D_d \, V_{\mathrm{CKM}}^\dagger + \kappa D_u$. 
With input data of six quark masses, three angles and one CP-phase in the 
CKM matrix and three charged-lepton masses, we can solve the above equations 
and determine $\kappa$ and $|c_d|$, but one parameter, the phase of $c_d$, 
is left undetermined \cite{Matsuda:2000zp, Matsuda:2001bg, Fukuyama:2002ch}. 
With input data of six quark masses, three angles and one CP-phase 
in the CKM matrix and three charged lepton masses, 
we solve the above equations and determine $\kappa$. 
The original basic mass matrices, $M_{10}$ and $M_{126}$, are described by 
\begin{eqnarray}
M_{10} 
&=& 
\frac{3+ |c_d| e^{i \sigma}}{4} 
V_{\mathrm{CKM}}^* \, D_d \, V_{\mathrm{CKM}}^\dagger
+ \frac{|c_d| e^{i \sigma} \kappa}{4} D_u, 
\label{M10}
\\
M_{126} &=& 
\frac{1- |c_d| e^{i \sigma}}{4} 
V_{\mathrm{CKM}}^* \, D_d \, V_{\mathrm{CKM}}^\dagger
-\frac{|c_d| e^{i \sigma} \kappa}{4} D_u, 
\label{M126} 
\end{eqnarray}
as the functions of $\sigma$, the phase of $c_d$, with the solutions 
$|c_d|$ and $\kappa$ determined by the GUT relation.  

Now let us solve the GUT relation and determine $|c_d|$ and $\kappa$. 
Since the GUT relation of Eq.~(\ref{GUTrelation}) is valid only at the GUT 
scale, we first evolve the data at the weak scale to the corresponding 
quantities at the GUT scale with given $\tan \beta$ according to the 
renormalization group equations (RGE's) and use them as input data at the 
GUT scale. Note that it is non-trivial to find the solution of the GUT 
relation since the number of the free parameters (fourteen) is almost the 
same as the number of inputs (thirteen). The solution of the GUT relation 
exists only if we take appropriate input parameters. Taking the experimental 
data at the $M_Z$ scale \cite{FK}, we get the following values 
for charged fermion masses and the CKM matrix at the GUT scale, $M_{\rm GUT}$ 
with $\tan \beta = 10$: 
\begin{eqnarray}
& & m_u = 0.000980 \; , \; \; m_c = 0.285 \; , \; \;  m_t = 113, 
\nonumber\\
& & m_d = 0.00135 \; , \; \; m_s = 0.0201 \; , \; \; m_b = 0.996, 
\nonumber\\ 
& & m_e = 0.000326 \; , \; \; m_\mu = 0.0687 \; , \; \; m_\tau = 1.17, 
\nonumber
\end{eqnarray}
and 
\begin{eqnarray}
V_{\mathrm{CKM}}(M_G) 
= \left( 
\begin{array}{ccc}
0.975 & 0.222 & - 0.000940 - 0.00289 i \\
-0.222 - 0.000129 i & 0.974 + 0.000124 i & 0.0347 \\ 
0.00864 - 0.00282 i & - 0.0337 - 0.000647 i & 0.999
\end{array} 
\right) \;
\nonumber
\end{eqnarray}
in the standard parameterisation. The signs of the input fermion masses 
have been chosen to be $(m_u, m_c, m_t) = (+, -, +)$ and 
$(m_d, m_s, m_b) = (-, -, +)$. 
By using these outputs at the GUT scale as input parameters, 
we can solve Eqs.~(\ref{cond1}) and (\ref{cond2}) and find a solution: 
\begin{eqnarray}
& \kappa = - 0.0103 + 0.000606 i \;, \nonumber\\ 
& |c_d| = 6.32  \; . & 
\end{eqnarray}
Once these parameters, $|c_d|$ and $\kappa$, are determined, we can describe 
all the fermion mass matrices as a functions of $\sigma$ from the mass matrix 
formulas of Eqs.~(\ref{massmatrix}), (\ref{M10}) and (\ref{M126}). 
Thus in the minimal SO(10) model we have almost unambiguous Dirac neutrino 
mass matrix $m_D$ and, therefore, we can obtain the informations on $M_D$ 
from the neutrino experiments via 
$M_{\nu}=(M_D^{-1} m_D)^{T} \mu_s (M_D^{-1}m_D)$ as in Eq.~(\ref{typeIII}). 

Now we proceed to the numerical calculation of $M_D$ from 
the well-confirmed neutrino oscillation data. The MNS mixing matrix $U$ 
in the standard parametrization is 
\be
U=
\left(
\begin{array}{ccc}
c_{13}c_{12} & c_{13}s_{12}e^{i\varphi_2} & s_{13}e^{i(\varphi_1-\delta)} \\
(-c_{23}s_{12}-s_{23}c_{12}s_{13} e^{i\delta})e^{-i\varphi_2}
&c_{23}c_{12}-s_{23}s_{12}s_{13} e^{i\delta} 
&s_{23}c_{13} e^{i(\varphi_1-\varphi_2)} \\
(s_{23}s_{12}-c_{23}c_{12}s_{13} e^{i\delta})e^{-i\varphi_1}
 & (-s_{23}c_{12}-c_{23}s_{12}s_{13} e^{i\delta})
e^{-i(\varphi_1-\varphi_2)} 
& c_{23}c_{13} \\
\end{array}
\right)\;,
\label{MNS}
\ee
where $s_{ij}:= \sin \theta_{ij}$, $c_{ij}:=\cos \theta_{ij}$ 
and $\delta$, $\varphi_1$, $\varphi_2$ are the Dirac phase and 
the Majorana phases, respectively. 
Recent KamLAND data tells us that 
\footnote{Our convention is $\Delta m_{ij}^2 = m_i^2 -m_j^2$.}
\bea
\Delta m^2_{\oplus} &=& 
\Delta m^2_{32} 
\ =\ 2.1 \times 10^{-3}\;\; {\mathrm{eV}}^2\;,
\nonumber\\
\sin^2 \theta_{\oplus} &=& 0.5\;,
\nonumber\\
\Delta m^2_{\odot} &=& 
\left| \Delta m^2_{21} \right| 
\ =\ 8.3 \times 10^{-5}\;\; {\mathrm{eV}}^2\;,
\nonumber\\
\sin^2 \theta_{\odot} &=& 0.28\;,
\nonumber\\
|U_{e3}|^2 &<& 0.061\;. \label{eqs030401}
\eea
For simplicity we take $U_{e3}=0$. Note that we can take both signs 
of $\Delta m^2_{21}$, $\Delta m^2_{21} > 0$ or $\Delta m^2_{21} < 0$.  
The former is called normal hierarchy, the latter is called inverted 
hierarchy. Here we adopt the former case, and take the lightest neutrino 
mass eigenvalue as $m_\ell = 10^{-3} \;{\rm [eV]}$. 
Then the mass eigenvalues are written as 
\bea
m_1 &=& m_\ell\;,
\nonumber\\
m_2 &=& \sqrt{m_\ell^2 + \Delta m^2_{\oplus}}\;,
\nonumber\\
m_3 &=& \sqrt{m_\ell^2 + \Delta m^2_{\oplus} + \Delta m^2_{\odot}}\;.
\eea
For the light Dirac neutrino mass matrix $m_D$, we input the SO(10) 
predicted one as was done in the previous section. However, unlike 
the case of minimal SO(10) GUT model, we can not fix $\sigma$. 
So we can obtain the heavy Dirac neutrino mass matrix $M_D$ as 
a function of $\mu_s$ and the three undetermined parameters, 
$\sigma$, two Majorana phases $\varphi_1$ and $\varphi_2$. 
For example, for fixed $\mu_s =1~{\rm [keV]}$ 
(For the implication of this value, see the remarks below Eq.~(\ref{MD}).) 
and $\varphi_1 = \varphi_2 = 0$, 
we get a prediction for the mass spectra of $M_D$. 
The dependences on the parameters $\sigma$ and $U_{e3}$ 
for fixed $\sigma = \pi$ are depicted in Fig.~\ref{Fig1}. 
These values are allowed by the present experiments \cite{Achard:2001qv} 
and are accessible and testable by the Large Hadron Collider (LHC) 
at CERN, in which we are able to discover new particles with masses up to 
$\lesssim 7$ [TeV] \cite{Gninenko:2003fv}. 
\FIGURE[t]{
\let\picnaturalsize=N
\def\picsize{7.5cm}
\ifx\nopictures Y\else{
\let\epsfloaded=Y
\centerline{\hspace{6mm}
{\ifx\picnaturalsize N\epsfxsize \picsize\fi
\epsfbox{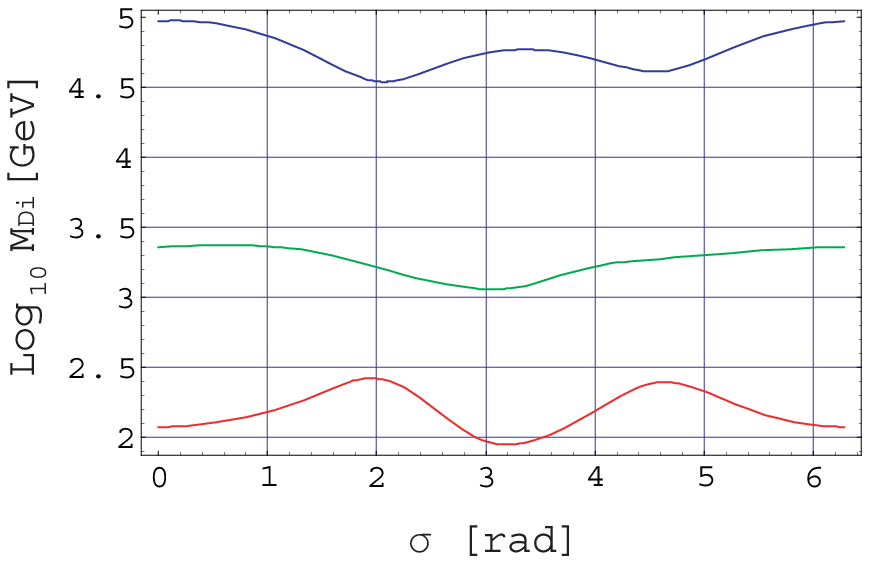}
{\ifx\picnaturalsize N\epsfxsize \picsize\fi}}}
\centerline{\hspace{6mm}
{\ifx\picnaturalsize N\epsfxsize \picsize\fi
\epsfbox{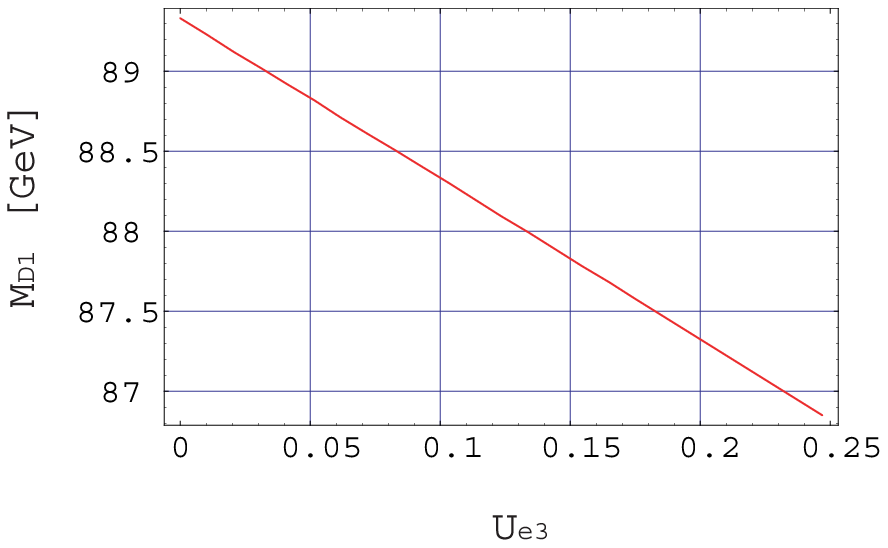}
{\ifx\picnaturalsize N\epsfxsize \picsize\fi
\epsfbox{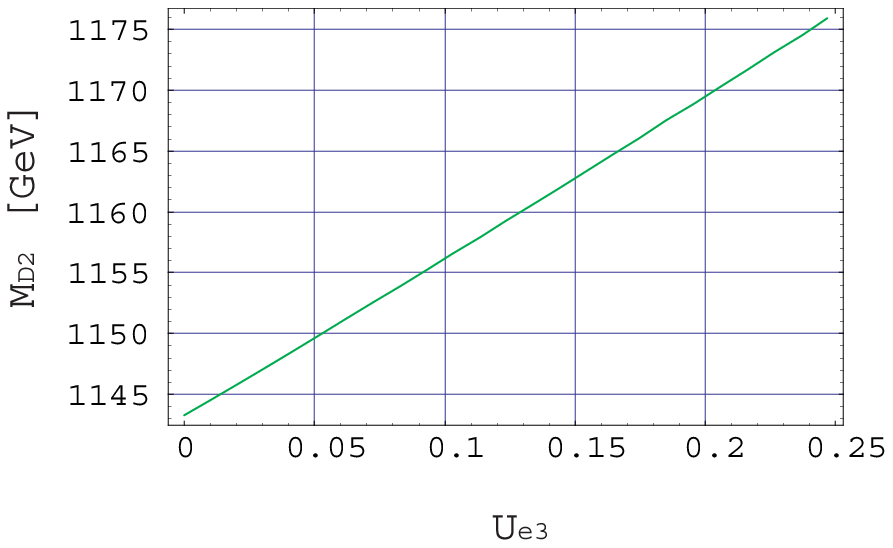}} } }
}\fi
\vspace{-6mm}
\caption{The predicted mass spectra of an additional singlet neutrino 
$M_{Di}~(i=1-3)$. The top panel represents three mass eigenvalues 
as a function of $\sigma$, the second and the third panels are 
the lightest and the second lightest masses as a function of $U_{e3}$ 
for fixed $\sigma=\pi$. \label{Fig1}}}

Of course, these values depend on the ambiguous assumptions taken above. 
We may take another strategy adopted in \cite{Matsuda:2004bq}. 
As shown in the paper \cite{Matsuda:2004bq}, 
we repeat the substitution of the normally-distributed random numbers 
which give the experimental values \cite{Eidelman:2004wy}:
\begin{eqnarray}
 \left| {m_u \left( {2~{\rm{GeV}}} \right)} \right| 
&=& 2.9 \pm 0.6~{\rm{[MeV]}}, \quad
 \left| {m_d \left( {2~{\rm{GeV}}} \right)} \right| 
= 5.2 \pm 0.9~{\rm{[MeV]}}, \label{eq011901} \\
 \left| {m_s \left( {2~{\rm{GeV}}} \right)} \right| 
&=& 99 \pm 16~{\rm{[MeV]}}, \quad
 \left| {m_c \left( {m_c} \right)} \right| 
= 1.0 - 1.4~{\rm{[GeV]}},\\
 m_b \left( {m_b} \right) 
&=& 4.0 - 4.5~{\rm{[GeV]}}, \quad
 m_t^{{\rm{direct}}}  = 174.3 \pm 5.1~{\rm{[GeV]}},
\end{eqnarray}
\begin{eqnarray}
 \left| {m_e^{{\rm{pole}}} } \right| 
&=& {\rm{0.510998902}} \pm {\rm{0.000000021~[MeV]}},\\ 
 \left| {m_\mu ^{{\rm{pole}}} } \right| 
&=& {\rm{105.658357}} \pm {\rm{0.00005,}} \quad
 m_\tau ^{{\rm{pole}}} = {\rm{1776.99}} \pm {\rm{0.29~[MeV]}},
\end{eqnarray}
\begin{eqnarray} 
 \sin \theta _{12} &=& 0.2229 \pm 0.0022, \quad
 \sin \theta _{23} = 0.0412 \pm 0.0020,\\
 \sin \theta _{13} &=& 0.0036 \pm 0.0007, \quad
 \delta  = \left( {59 \pm 13} \right)^\circ \label{eq011902}
\end{eqnarray}
for the quark and charged lepton masses and the CKM mixing and 
the Dirac phase parameters 10,000 times. 
On the other hand, about the remaining parameters,  
we assume Eq.~(\ref{eqs030401}), \(m_1 = 10^{-3}\) [eV] 
and \(U_{e3} = 0\) at the GUT scale, 
and Majorana phases and \(\sigma\) move from $0$ to $2 \pi$ 
in 8 equal intervals. 
Namely, we scan the possible ranges of undetermined parameters $\sigma$, 
$\varphi_1$, $\varphi_2$ and plotted the three masses of $M_{D}$, 
the three mixing angles 
and five phases of ${\cal U}$ which diagonalises the mass matrix $M_D$ 
in the basis where \(M_e\) is real diagonal in Figs.~\ref{Fig2}--\ref{Fig5}. 
Here we calculated the distributions for sixteen sets of possible 
combinations of mass signatures of up-type and down-type quarks. 
Figure \ref{Fig1} corresponds to the blue solid line 
of Figure \ref{Fig2} with $\mu_s =1$ [KeV]. 
\FIGURE[t]{\epsfig{file=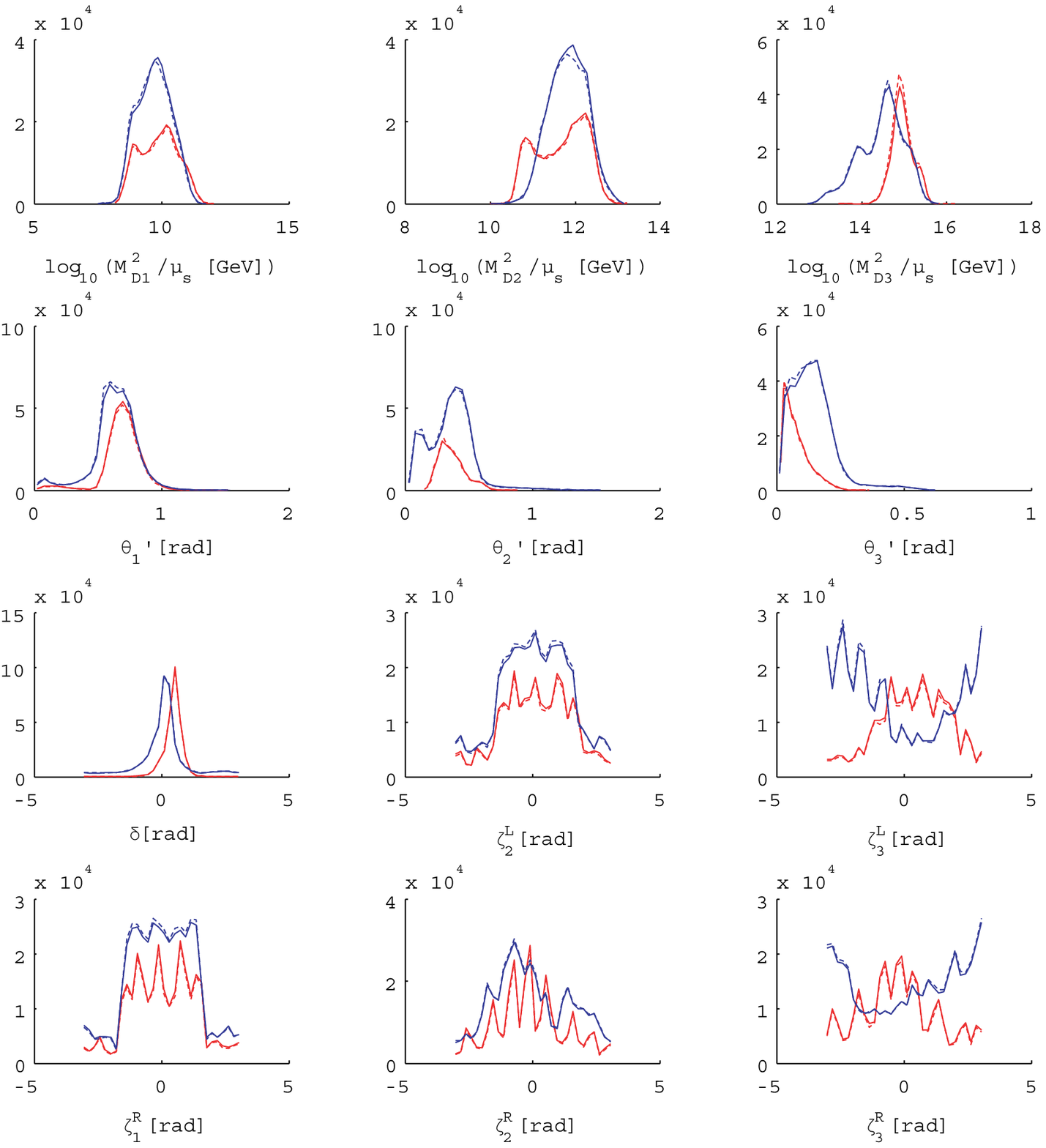, width=.6\textwidth}
\caption{The distributions of the predicted mass ratios, 
mixing angles and phases for $M_D$. 
The signs of each mass eigenvalues are chosen as follows: 
The red solid line is $(m_u,m_c,m_t)=(+,+,+);~(m_d,m_s,m_b)=(+,+,+)$, 
the red dotted one is $(m_u,m_c,m_t) = (-,+,+);~(m_d,m_s,m_b) = (+,+,+)$, 
the blue solid one is $(m_u,m_c,m_t)=(+,-,+);~(m_d,m_s,m_b)=(-,-,+)$ and 
the blue dotted one is $(m_u,m_c,m_t)=(-,-,+);~(m_d,m_s,m_b)=(-,-,+)$. 
\label{Fig2}}}

\FIGURE[t]{\epsfig{file=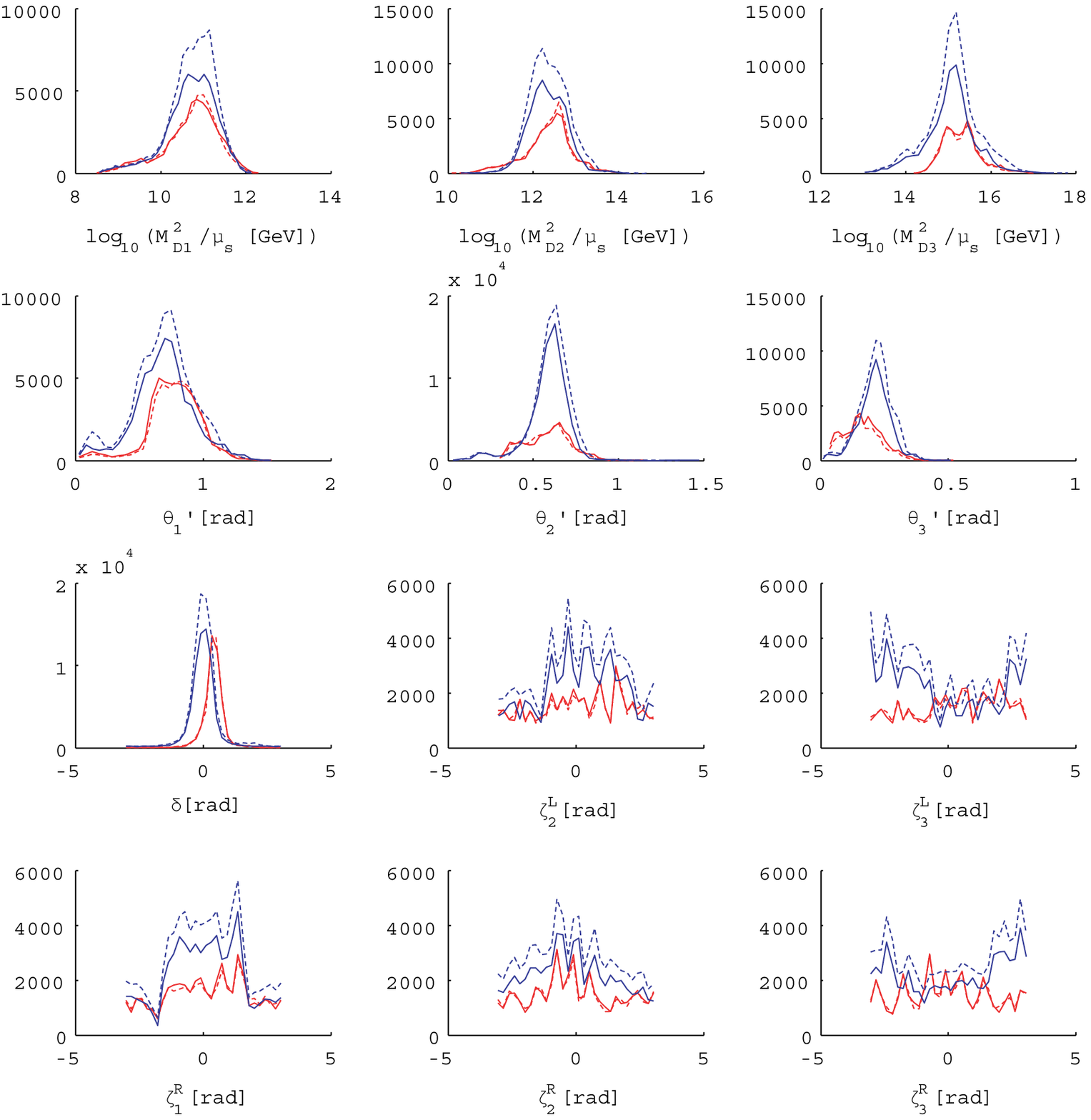, width=.6\textwidth}
\caption{The same distributions as Figure \ref{Fig2} but 
the signs of each mass eigenvalues are chosen as follows: 
The red solid line is $(m_u,m_c,m_t)=(+,-,+);~(m_d,m_s,m_b)=(+,+,+)$, 
the red dotted one is $(m_u,m_c,m_t)=(-,-,+);~(m_d,m_s,m_b)=(+,+,+)$, 
the blue solid one is $(m_u,m_c,m_t)=(+,+,+);~(m_d,m_s,m_b)=(-,-,+)$ and 
the blue dotted one is $(m_u,m_c,m_t)=(-,+,+);~(m_d,m_s,m_b)=(-,-,+)$. 
\label{Fig3}}}

\FIGURE[t]{\epsfig{file=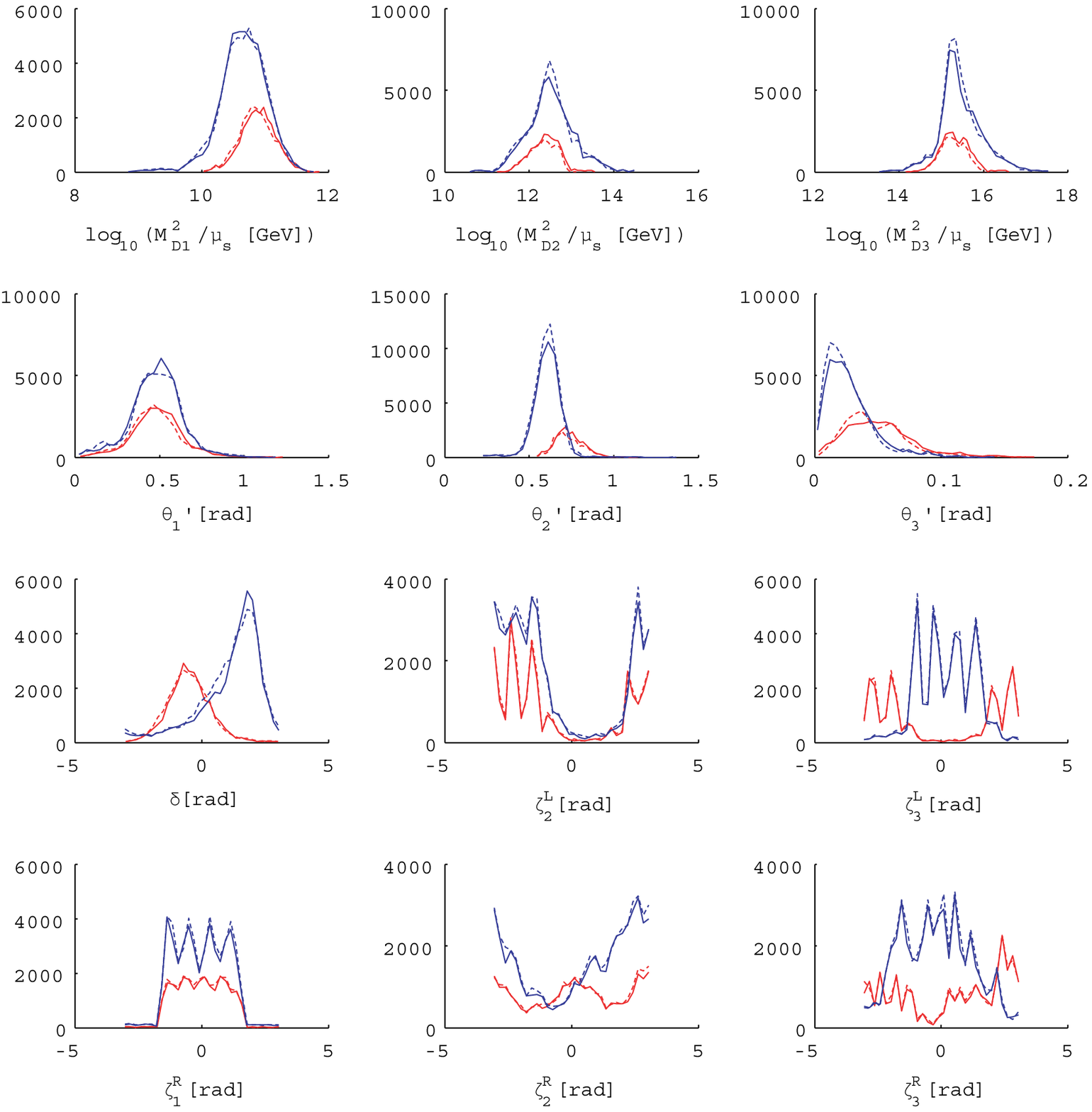, width=.6\textwidth}
\caption{The same distributions as Figure \ref{Fig2} but 
the signs of each mass eigenvalues are chosen as follows: 
The red solid line is $(m_u,m_c,m_t)=(+,+,+);~(m_d,m_s,m_b)=(-,+,+)$, 
the red dotted one is $(m_u,m_c,m_t)=(-,+,+);~(m_d,m_s,m_b)=(-,+,+)$, 
the blue solid one is $(m_u,m_c,m_t)=(+,-,+);~(m_d,m_s,m_b)=(+,-,+)$ and 
the blue dotted one is $(m_u,m_c,m_t)=(-,-,+);~(m_d,m_s,m_b)=(+,-,+)$. 
\label{Fig4}}}

\FIGURE[t]{\epsfig{file=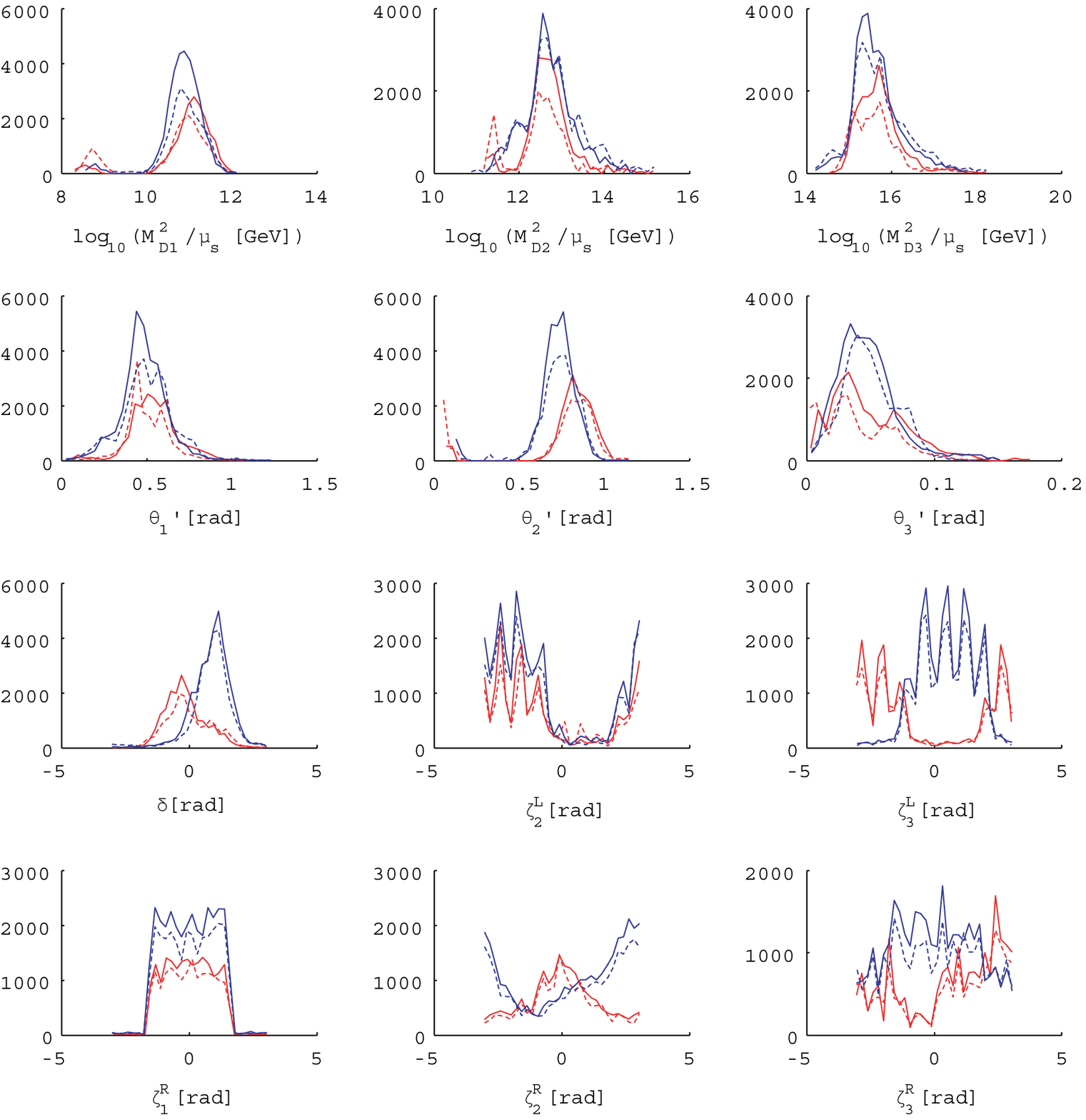, width=.6\textwidth}
\caption{The same distributions as Figure \ref{Fig2} but 
the signs of each mass eigenvalues are chosen as follows: 
The red solid line is $(m_u,m_c,m_t)=(+,-,+);~(m_d,m_s,m_b)=(-,+,+)$, 
the red dotted one is $(m_u,m_c,m_t)=(-,-,+);~(m_d,m_s,m_b)=(-,+,+)$, 
the blue solid one is $(m_u,m_c,m_t)=(+,+,+);~(m_d,m_s,m_b)=(+,-,+)$ and 
the blue dotted one is $(m_u,m_c,m_t)=(-,+,+);~(m_d,m_s,m_b)=(+,-,+)$. 
\label{Fig5}}}

Finally, it is remarkable to say that the see-saw mechanism itself 
(or the types of it) can never been proofed and all the models should 
take care of all the types of the see-saw mechanism including 
the alternatives to it \cite{Murayama:2004me, Smirnov:2004hs}. 
The test of all these models is due to the applications to the other 
phenomelogical consequences, for example, the lepton flavour violating 
processes and so on \cite{Deppisch:2004fa, Ilakovac:1994kj}. 

\section{Summary}
In this paper, we have constructed an SO(10) model in which the smallness 
of the neutrino masses are explained in terms of the type-III see-saw 
mechanism. To evaluate the parameters related to the singlet neutrinos, 
we have used the minimal SUSY SO(10) model. This model can simultaneously 
accommodate all the observed quark-lepton mass matrix data with appropriately 
fixed free parameters. Especially, the neutrino-Dirac-Yukawa coupling 
matrix are completely determined. Using this Yukawa coupling matrix, 
we have calculated the masses and mixings for the not-so-heavy singlet 
neutrinos. The obtained ranges of the mass of $M_D$ is interesting since 
they are testable by a forthcoming LHC experiment. 

\acknowledgments{
The work of T.F. is supported in part by the Grant-in-Aid for 
Scientific Research from the Ministry of Education, Science and Culture 
of Japan (\#16540269). He is also grateful to Professors D. Chang and 
K. Cheung for their hospitality at NCTS. The work of T.K. and K.M. 
are supported by the Research Fellowship of the Japan Society 
for the Promotion of Science (\#7336 and \#3700). 
The work of A.I. is supported by the Ministry of Science and 
Technology of Republic of Croatia under contract (\#0119261). }


\begin{thebibliography}{99}
\bibitem{Weinberg:1979sa}
S.~Weinberg,
Phys.\ Rev.\ Lett.\  {\bf 43}, 1566 (1979).

\bibitem{see-saw}
T.~Yanagida, in Proceedings of the workshop 
on the Unified Theory and Baryon Number in the Universe, 
edited by O.~Sawada and A.~Sugamoto (KEK, Tsukuba, 1979);

M.~Gell-Mann, P.~Ramond, and R.~Slansky, 
in Supergravity, edited by D.~Freedman and P.~van~Niewenhuizen 
(north-Holland, Amsterdam 1979); 

R.~N.~Mohapatra and G.~Senjanovi\'c, 
Phys.\ Rev.\ Lett.\ {\bf 44}, 912 (1980).

\bibitem{Zeller:2001hh}
G.~P.~Zeller {\it et al.}  [NuTeV Collaboration],
Phys.\ Rev.\ Lett.\  {\bf 88}, 091802 (2002)
[Erratum-ibid.\  {\bf 90}, 239902 (2003)]
[arXiv:hep-ex/0110059];
The theoretical explanations of this anomaly by including 
heavy singlet neutrinos are given in \cite{Takeuchi:2002nn}. 

\bibitem{Takeuchi:2002nn}
T.~Takeuchi,
[arXiv:hep-ph/0209109];

W.~Loinaz, N.~Okamura, T.~Takeuchi and L.~C.~R.~Wijewardhana,
Phys.\ Rev.\ D {\bf 67}, 073012 (2003)
[arXiv:hep-ph/0210193];

W.~Loinaz, N.~Okamura, S.~Rayyan, T.~Takeuchi and L.~C.~R.~Wijewardhana,
Phys.\ Rev.\ D {\bf 70}, 113004 (2004)
[arXiv:hep-ph/0403306].

\bibitem{EW86}
E.~Witten,
Nucl.\ Phys.\ B {\bf 268}, 79 (1986).

\bibitem{Mohapatra:1986aw}
R.~N.~Mohapatra,
Phys.\ Rev.\ Lett.\  {\bf 56} (1986) 561.

\bibitem{Mohapatra:1986bd}
R.~N.~Mohapatra and J.~W.~F.~Valle,
Phys.\ Rev.\ D {\bf 34}, 1642 (1986).

\bibitem{Val86}
J.~W.~F.~Valle, in 
NUCLEAR BETA DECAYS AND NEUTRINO: proceedings. 
Edited by T.~Kotani, H.~Ejiri, E.~Takasugi. 
Singapore, World Scientific, 1986. 542p. 

\bibitem{Barr:2003nn}
S.~M.~Barr,
Phys.\ Rev.\ Lett.\  {\bf 92}, 101601 (2004)
[arXiv:hep-ph/0309152].

\bibitem{Melo:1996ht}
I.~Melo,
[arXiv:hep-ph/9612488].

\bibitem{Berezhiani:1992cd}
Z.~G.~Berezhiani, A.~Y.~Smirnov and J.~W.~F.~Valle,
Phys.\ Lett.\ B {\bf 291}, 99 (1992)
[arXiv:hep-ph/9207209].

\bibitem{Berezinsky:1993fm}
V.~Berezinsky and J.~W.~F.~Valle,
Phys.\ Lett.\ B {\bf 318}, 360 (1993)
[arXiv:hep-ph/9309214].

\bibitem{Babu:1992ia}
K.~S.~Babu and R.~N.~Mohapatra,
Phys.\ Rev.\ Lett.\  {\bf 70}, 2845 (1993)
[arXiv:hep-ph/9209215].

\bibitem{Matsuda:2000zp}
K.~Matsuda, Y.~Koide and T.~Fukuyama,
Phys.\ Rev.\ D {\bf 64}, 053015 (2001)
[arXiv:hep-ph/0010026].

\bibitem{Matsuda:2001bg}
K.~Matsuda, Y.~Koide, T.~Fukuyama and H.~Nishiura,
Phys.\ Rev.\ D {\bf 65}, 033008 (2002)
[Erratum-ibid.\ D {\bf 65}, 079904 (2002)]
[arXiv:hep-ph/0108202];

\bibitem{Fukuyama:2002ch}
T.~Fukuyama and N.~Okada,
JHEP {\bf 0211}, 011 (2002)
[arXiv:hep-ph/0205066].

\bibitem{Goh:2003sy}
H.~S.~Goh, R.~N.~Mohapatra and S.~P.~Ng,
Phys.\ Lett.\ B {\bf 570}, 215 (2003)
[arXiv:hep-ph/0303055].

\bibitem{Goh:2003hf}
H.~S.~Goh, R.~N.~Mohapatra and S.~P.~Ng,
Phys.\ Rev.\ D {\bf 68}, 115008 (2003)
[arXiv:hep-ph/0308197].

\bibitem{Bajc:2002iw}
B.~Bajc, G.~Senjanovi\'c and F.~Vissani,
Phys.\ Rev.\ Lett.\  {\bf 90}, 051802 (2003)
[arXiv:hep-ph/0210207].

\bibitem{Dutta:2004wv}
B.~Dutta, Y.~Mimura and R.~N.~Mohapatra,
Phys.\ Rev.\ D {\bf 69}, 115014 (2004)
[arXiv:hep-ph/0402113].

\bibitem{Matsuda:2004bq}
K.~Matsuda,
Phys.\ Rev.\ D {\bf 69}, 113006 (2004)
[arXiv:hep-ph/0401154].

\bibitem{Matsuda:1999yx}
K.~Matsuda, T.~Fukuyama and H.~Nishiura,
Phys.\ Rev.\ D {\bf 61}, 053001 (2000)
[arXiv:hep-ph/9906433].

\bibitem{FK}
H.~Fusaoka and Y.~Koide,
Phys.\ Rev.\ D {\bf 57}, 3986 (1998)
[arXiv:hep-ph/9712201].

\bibitem{Achard:2001qv}
P.~Achard {\it et al.}  [L3 Collaboration],
Phys.\ Lett.\ B {\bf 517}, 67 (2001)
[arXiv:hep-ex/0107014].

\bibitem{Gninenko:2003fv}
S.~N.~Gninenko, M.~M.~Kirsanov, N.~V.~Krasnikov and V.~A.~Matveev,
[arXiv:hep-ph/0301140].

\bibitem{Eidelman:2004wy}
S.~Eidelman {\it et al.}  [Particle Data Group],
Phys.\ Lett.\ B {\bf 592}, 1 (2004).

\bibitem{Murayama:2004me}
H.~Murayama,
Nucl.\ Phys.\ Proc.\ Suppl.\  {\bf 137}, 206 (2004)
[arXiv:hep-ph/0410140].

\bibitem{Smirnov:2004hs}
A.~Y.~Smirnov,
[arXiv:hep-ph/0411194].

\bibitem{Deppisch:2004fa}
F.~Deppisch and J.~W.~F.~Valle,
[arXiv:hep-ph/0406040].

\bibitem{Ilakovac:1994kj}
A.~Ilakovac and A.~Pilaftsis,
Nucl.\ Phys.\ B {\bf 437}, 491 (1995)
[arXiv:hep-ph/9403398].
\end{thebibliography}
\end{document}